\documentclass[12pt]{article}
\usepackage{a4wide}
\usepackage{amsmath}
\usepackage{amssymb}
\usepackage{graphicx}
\usepackage{latexsym,epsfig,cite}
\usepackage{booktabs}
\usepackage{hyperref}

\def\beq{\begin{equation}}
\def\eeq{\end{equation}}
\def\bea{\begin{eqnarray}}
\def\eea{\end{eqnarray}}

\def\lp#1{{\lambda^{\prime}_{#1}}}
\def\lpp#1{{\lambda^{\prime\prime}_{#1}}}

\def\lsim{\mathrel{\rlap{\raise 2.5pt \hbox{$<$}}\lower 2.5pt\hbox{$\sim$}}}
\def\gsim{\mathrel{\rlap{\raise 2.5pt \hbox{$>$}}\lower 2.5pt\hbox{$\sim$}}}

\allowdisplaybreaks 

\usepackage[dvips]{color}
\definecolor{Black}{named}{Black}
\definecolor{Red}{named}{Red}

\begin{document}

\thispagestyle{empty}
\begin{flushright}

\end{flushright}
\vspace*{5mm}
\begin{center}
{\large {\bf R-parity violating chargino decays at the LHC
}}\\
\vspace*{1cm}
{\bf N.-E.\ Bomark$^1$, A.\ Kvellestad$^2$, S.\ Lola$^3$,}
{\bf P.\ Osland$^4$ and A.R.\ Raklev$^{2,5}$} \\
\vspace{0.3cm}
$^1$ National Centre for Nuclear Research, Hoza 69, 00-681 Warsaw, Poland.\\
$^2$ Department of Physics, University of Oslo, N-0316 Oslo, Norway\\
$^3$ Department of Physics, University of Patras, GR-26500 Patras, Greece \\
$^4$ Department of Physics and Technology, University of Bergen,
N-5020 Bergen, Norway \\
$^5$ Department of Applied Mathematics and Theoretical Physics, University of Cambridge,
Wilberforce Road, Cambridge CB3 0WA, UK
\end{center}

\begin{abstract}
Supersymmetric models with R-parity violation  (RPV) have become more
popular following the lack of any excess of missing energy events
at the 8 TeV LHC. To identify such models, the suggested searches
generally rely on the decay products of the (effectively) lightest
supersymmetric particle (LSP), with signals that depend on the
identity of the LSP and the relevant RPV operators. Here we
look at the prospects for detecting RPV chargino decays at the LHC and find substantial patches of parameter space in the Minimal Supersymmetric Standard Model with possibly spectacular signatures, such as three charged-lepton resonances.
\end{abstract}

\setcounter{page}{1}

\section{Introduction}

Most searches for supersymmetry with R-parity violation (RPV), see {\it e.g.}\ \cite{Barbier:2004ez} for a review, adopt
one of the following approaches: either one looks for the effects of the
relevant operators at low energies in precision measurements, or one
focuses on the decay of the (effectively) lightest supersymmetric particle (LSP)
through RPV operators. This is the result of the strict
bounds on most such couplings, which imply that RPV decay widths are
typically subdominant, unless they are the only viable option. For the same reason, possibly spectacular
signatures, such as resonant single superparticle productions, are
 only viable for a limited number of operators and for a
constrained range of couplings. In most cases, the LSP,
with no other alternative decays available, is the best candidate to look at.

The plethora of RPV operators implies that there is a broad range of potential
signals \cite{Barbier:2004ez}, which
also depend on the nature of the LSP. In the presence of RPV, any
sparticle may in principle be the LSP, since bounds on stable massive
charged particles no longer apply. However, the lightest chargino is not
usually considered, because common lore
says that  the structure of the gaugino mass matrices is such that
the lightest neutralino is always lighter than the lightest chargino.

In this paper we point out that this is not necessarily true. As
discussed by Kribs {\it et al.}\ \cite{Kribs:2008hq}, in the general
Minimal Supersymmetric Standard Model (MSSM) the chargino could in
principle be lighter than the lightest neutralino in a corner of
the parameter space. Even when heavier, the small mass difference that
is possible and even natural in certain scenarios, implies
that the chargino, as the next-to-lightest sparticle (NLSP), can have
dominant decays through RPV operators.

Models with almost-degenerate electroweak gauginos, where such chargino decays
can be expected, arise for example
in the context of anomaly-mediated supersymmetry breaking ~\cite{Randall:1998uk,Giudice:1998xp}. The current absence of any
supersymmetry (SUSY) signal at the LHC, and the question of the naturalness of the
remaining parameter space, has led to the
consideration of so-called Natural SUSY models. Here only the
higgsinos, the stops, the left-handed sbottom, and, to a more limited
extent, the gluino, are light enough to be probed at the
LHC~\cite{Brust:2011tb,Papucci:2011wy}. Higgsino dominance of
light neutralinos and charginos also leads to small mass differences, although we have recently shown that, given current direct and indirect constraints, the degeneracy is considerably less severe than for winos~\cite{Bomark:2013nya}.

In R-parity conserving (RPC) models, if the degeneracy is severe
enough, such spectra can lead to rather characteristic experimental
signals in high-energy collisions, with charginos that live long
enough to create displaced vertices, or even pass through the detector
before decaying
\cite{Chen:1996ap,Feng:1999fu,Gherghetta:1999sw}.
Here we instead study the consequences
of lifting the RPC. In particular, we are here interested in the trilinear RPV operators in the superpotential:
\begin{equation}
W \sim\lambda_{ijk} L_iL_j\bar{E}_k+\lp{ijk}L_iQ_j\bar{D}_k+\lpp{ijk}\bar{U}_i\bar{D}_j\bar{D}_k.
\end{equation}
We do not include bilinear RPV operators of the form $\mu_iH_uL_i$. These will induce mixing between the charginos and the charged leptons, and allow decays for the lightest chargino of the form $\tilde\chi^\pm_1\to \nu_iW^\pm,l_i^\pm Z$. These decays have already been discussed for AMSB scenarios motivated by neutrino masses in \cite{deCampos:2008av}.

Direct chargino decays
via subsets of RPV operators have been considered in the past,
especially in the context of LEP physics, see {\it
  e.g.}~\cite{Dreiner:1996dd}.
Nevertheless, a detailed discussion that also takes into account the
rich flavour structure of the RPV operators is still
lacking. Here, we investigate the consequences of direct RPV chargino
decays in the context of LHC searches, and we show that this can
lead to dramatic signals, such as resonant three-lepton final
states due to $LL\bar E$ operators. In addition, the presence of
heavy quarks in $LQ\bar{D}$ and $\bar{U}\bar{D}\bar{D}$ operators
results in enhanced detection prospects in this case as well.

One may object to breaking R-parity, since it ensures the existence of
a stable sparticle, realising one of the central motivations for
weak-scale supersymmetry: the existence of dark matter. However, as
pointed out in recent years \cite{TY, BM, LOR}, the gravitino may be
the {\it real} LSP, with a naturally long lifetime, due to its tiny
gravitational coupling. For a wide range of parameters gravitinos are
essentially stable on cosmological time-scales and can act as dark
matter. Despite this, we will refer to the neutralino and chargino as
the {\it effective} LSP to avoid a very convoluted language.\footnote{Note also that this is not compatible with the standard
anomaly-mediated breaking scenario, where the gravitino is heavy
compared to the other sparticles, which have loop-suppressed masses.}

In this paper, we present an update on how near-degenerate electroweak gauginos arise in
general supersymmetry breaking models described by the MSSM, and to what extent they are compatible with the recent discovery of a new boson at the LHC, when interpreted as the light SM-like Higgs state of the MSSM~\cite{ATLAS:2012gk,CMS:2012gu}, and other direct and indirect constraints.  We study the possibility of a light chargino with the
 {\tt MultiNest~2.17} \cite{Feroz:2007kg,Feroz:2008xx} code for parameter sampling, which enables a detailed analysis of the posterior probability distribution in the supersymmetric parameter space of interest, taking into account the available experimental constraints.
We then study the effect of RPV operators on these models in the
context of searches at the LHC.

We should note that the interesting properties of chargino decays that
we find within our scenario,
also apply to extensions of the theory
beyond the MSSM. In this respect direct chargino decays can
be a powerful tool to probe the gaugino sector and distinguish
between different possibilities.  Significant
deviations from gaugino unification can arise naturally in
well-motivated scenarios, for instance due to the presence of F-terms
\cite{Ellis:1985jn,Martin:2009ad}; in such schemes, the specific
gaugino hierarchies to be expected are
fixed to a large extent by the group theory, and the argument can be reversed:
if direct chargino decays via RPV operators are detected at
significant rates, we will have crucial information
on the structure of the gaugino sector and on the underlying GUT symmetries of the theory.

We begin in Section \ref{sec:massdiff} by discussing the parameters
that affect the neutralino--chargino mass difference in the MSSM. We
then describe the parameter scan that we have performed in
Section~\ref{sec:scan}. In Section~\ref{sec:current} we look at the consequences of current bounds on the RPV couplings and competition with the RPC
chargino decays. In Section~\ref{sec:vertex} we disucss the impact of current LHC searches for displaced vertices on our scenario.  Finally, we describe the consequences of our results for LHC searches for RPV chargino decays in Section~\ref{sec:lhc}, before we conclude in Section~\ref{sec:conclusions}.

\section{Neutralino--chargino mass difference}
\label{sec:massdiff}
In the MSSM, the free mass parameters in the neutralino mass matrix at tree level are $M_1$, $M_2$ and $\mu$. In addition, $\tan\beta$ also enters as a free parameter. With the exception of $M_1$, the same set of parameters enters in the tree-level chargino mass matrix. Any complex phases for the mass parameters are very constrained, in particular due to limits on the electric dipole moments \cite{Feng:2008cn,Ellis:2008zy,Cheung:2009fc,Altmannshofer:2009ne}. However, there are no {\it a priori} grounds not to give arbitrary signs to these parameters, although by a rotation of basis we can choose $M_2$ to always be positive.

For small $M_1$ the lightest neutralino will be a bino, which is
historically the most popular choice. Since $M_1$ does not enter in the
chargino mass matrix, in this case there is no degeneracy
between chargino and neutralino.  When $\mu$ or $M_2$ is the smallest parameter
we may have a neutralino that is dominantly a higgsino or a wino,
and in both cases there may be degeneracy with the chargino.

In the wino limit, $M_2<|M_1|,\mu$, the tree-level mass difference
\begin{equation}
\Delta m\equiv m_{\tilde\chi_1^{\pm}}-m_{\tilde\chi_1^0},
\end{equation}
expanded in $1/\mu$ is \cite{Feng:1999fu,Gherghetta:1999sw}
\begin{eqnarray}
\Delta m &=& \frac{M_W^2}{\mu^2}\frac{M_W^2}{M_1-M_2}\tan^2\theta_W\sin^2{2\beta}+2\frac{M_W^4M_2\sin{2\beta}}{(M_1-M_2)\mu^3}\tan^2\theta_W\nonumber\\
&&+\frac{M_W^6\sin^3{2\beta}}{(M_1-M_2)^2\mu^3}\tan^2\theta_W(\tan^2\theta_W-1)+{\mathcal O}\left(\frac{1}{\mu^4}\right).
\label{eq:dmwinotree}
\end{eqnarray}
Note that while this can give a negative $\Delta m$ for negative $M_1$, these tree-level terms are all small for large $\tan\beta$, and for $\tan\beta\to\infty$ the lowest contributing order is in fact $1/\mu^4$. This means that loop effects can be significant. The leading loop correction from gauge bosons---assuming there are no very light sfermions---is positive and in the wino limit is given by \cite{Cheng:1998hc,Feng:1999fu}
\begin{equation}
\Delta m_{\rm 1-loop}= \frac{\alpha_2M_2}{4\pi}[f(M_W/M_2)-\cos^2\theta_Wf(M_Z/M_2)-\sin^2\theta_Wf(0)],
\label{eq:dmwino1loop}
\end{equation}
where
\begin{equation}
f(a)=2\int_0^1(1+x)\ln{(x^2+(1-x)a^2)}\,dx.
\end{equation}
In the limit where $M_2\gg M_W$, this gives $\Delta m_{\rm
  1-loop}\simeq 165$ MeV.\footnote{Note that, apart from a numerical
  factor, this electroweak correction is, as expected, $\alpha M_W$, which just happens to be of the order of the pion mass.} The possibility of getting a mass difference $\lsim 165$ MeV then rests on the contribution from Eq.~(\ref{eq:dmwinotree}) being negative and significant compared to the expression (\ref{eq:dmwino1loop}). This could be the case for negative $M_1$.

In the higgsino limit, $|M_1|,M_2>\mu,M_W$, the tree-level mass difference from an expansion in $1/M_2$ is \cite{Giudice:1995qk}:
\begin{equation}
\Delta m = \left[\frac{M_2}{M_1}\tan^2\theta_W+1+{\rm sgn}\,\mu\left(\frac{M_2}{M_1}\tan^2\theta_W-1\right)\sin{2\beta}\right]\frac{M_W^2}{2M_2}+{\mathcal O}\left(\frac{1}{M_2^2}\right).\label{eq:dmhiggsino}
\end{equation}
This expansion breaks down for $\mu\to 0$, however, LEP limits on the chargino mass ensure that we can keep out of that region of parameter space.

For positive $M_1$ and $M_2$, $\Delta m$ in the higgsino limit is always
positive. It becomes small for very large $M_1,M_2\gg M_W$, but
numerically this does not lower the mass difference below 300 MeV,
unless (i) both masses are greater than ${\mathcal O}(10~{\rm TeV})$,
or (ii) $\tan\beta\simeq 1$ and either mass is very large.
For a negative $M_1$, however, we may have a negative $\Delta m$, but
this occurs for very special choices of the parameters, namely: relatively
small $|M_1|$, combined with large $M_2$ and small $\tan\beta$, see~\cite{Bomark:2013nya}. In
addition to the above, we have loop corrections that mainly stem from top--stop and
$\gamma(Z)$--higgsino loops. The former can have either sign depending
on the stop mixing, while the latter is small unless $\tan\beta$ is
large. Both are included in the scan that will be performed in the next Section.

\section{Parameter scan}
\label{sec:scan}
\subsection{Scan set-up}

To search for these degenerate models, we employ a bayesian scan over the MSSM parameter space, using the three parameters $M_1$, $M_2$ and $M_3$ to represent the gaugino mass for the $U(1)$, $SU(2)$ and $SU(3)$ sectors, respectively, at the electroweak scale. $M_1$ and $M_3$ are allowed to take negative values. For the Higgs sector we use the parameters $\mu$, $m_{A^0}$ and $\tan\beta$, deriving $m_{H_u}$ and $m_{H_d}$ from EWSB. These are the relevant parameters for the problem at hand. Furthermore, we use a common mass parameter $m_{\tilde{q}}$ for the first and second generation squarks, while a separate parameter $m_{\tilde{q}_3}$ is used for squarks of the third generation. Finally, the sleptons are governed by a common mass scale $m_{\tilde{l}}$, and we use a common value $A_0$ for the trilinear couplings. We do not scan over the RPV couplings directly, as the sheer number of couplings makes this unfeasible. However, in Section~\ref{sec:imp} we discuss in detail the allowed values of these couplings, and study how they affect the chargino lifetime and branching fractions for the set of posterior samples produced by the scan.

For all mass parameters we use logarithmic priors in order to incorporate a prior belief in naturalness~\cite{Cabrera:2008tj}, while for $\tan\beta$ a flat prior is used. The SM parameters $m_t$, $m_b$, $M_Z$, $\alpha$ and $\alpha_s$ are included as nuisance parameters with gaussian priors. A summary of the parameters used, along with ranges and priors, is given in Table~\ref{tab:ScanPar}. For comparison, we also perform a scan with flat priors for all parameters except the SM nuisance parameters. This scan is performed with less statistics compared to the main scan, as it is only used to check the prior dependence of our results.

\begin{table}[h!]
  \begin{center}
    \begin{small}
      \begin{tabular}{lccc}
        \toprule
  Parameter          &  Range                  &  Prior      &  Reference \\
        \midrule
  $M_1$              &  $[-4000,4000]$         &  log        &   -    \\
  $M_2$              &  $[0,4000]$             &  log        &   -    \\
  $M_3$              &  $[-4000,4000]$         &  log        &   -    \\
  $\mu$              &  $[-4000,4000]$         &  log        &   -    \\
  $m_{A^0}$          &  $[0,4000]$             &  log        &   -    \\
  $m_{\tilde{l}}$    &  $[0,7000]$             &  log        &   -    \\
  $m_{\tilde{q}}$    &  $[0,7000]$             &  log        &   -    \\
  $m_{\tilde{q}_3}$  &  $[0,7000]$             &  log        &   -    \\
  $A_0$              &  $[-7000,7000]$         &  log        &   -    \\
  $\tan\beta$        &  $[2,60]$               &  linear     &   -    \\
        \midrule
  $m_t$              & $173.4 \pm 1.0$         &  gaussian   & \cite{CMS:2012fya}  \\
  $m_b^{\overline{MS}}(m_b)$              & $4.18  \pm 0.03$        &  gaussian   & \cite{Beringer:1900zz}  \\
  $M_Z$              & $91.1876 \pm 0.0021$    &  gaussian   & \cite{Beringer:1900zz}  \\
  $\alpha^{-1}$      & $127.944 \pm 0.014$     &  gaussian   & \cite{Beringer:1900zz} \\
  $\alpha_s$         & $0.1184 \pm 0.0007$     &  gaussian   & \cite{Beringer:1900zz} \\
      \bottomrule
      \end{tabular}
    \end{small}
\caption{List of scan parameters with ranges and priors. Dimensionful
  parameters are given in GeV. All non-SM parameters are given at the
  scale $Q = 1.0$\;TeV, except $\tan\beta$ and $\mu$ which are
given at the EWSB scale, and the pseudoscalar Higgs pole mass $m_{A^0}$. Log priors are set to zero over the ranges $(-25,25)$ and $(0,25)$ GeV for signed and non-negative parameters, respectively.}\label{tab:ScanPar}
  \end{center}
\end{table}

The scan uses {\tt MultiNest 2.17} \cite{Feroz:2007kg,Feroz:2008xx} to
explore the parameter space described above. For each point in the parameter space,
the sparticle spectrum is calculated by {\tt SoftSusy
  3.3.5}~\cite{softsusy}, including the effects discussed in Section
\ref{sec:massdiff} for the chargino--neutralino mass difference, while
Higgs masses are calculated using {\tt FeynHiggs 2.9.4}~\cite{feynhiggs1,feynhiggs2,feynhiggs3,feynhiggs4}.
We apply constraints from electroweak precision observables and
B-physics, using {\tt SoftSusy} and {\tt MicrOMEGAS
  2.4.5}~\cite{micromegas1,micromegas2,micromegas3} to calculate the
relevant quantities. In addition, the relevant constraints from LEP data on the
chargino mass and the LHC Higgs mass measurement are included. We note that most limits on the chargino mass from LEP are void due to the small $\Delta m$.
The values and distributions used for these constraints are summarised in Table~\ref{tab:Constraints}. Note that although the CMS limit on BR$(B_s \rightarrow \mu\mu)$~\cite{Chatrchyan:2013bka}, is slightly more constraining than the corresponding LHCb limit~\cite{Aaij:2013aka,Aaij:2013aka_sup}, since the latter provides a likelihood covering a wider range of branching ratio values, we use that in our scan.

\begin{table}[h!]
  \begin{center}
    \begin{small}
      \begin{tabular}{lccc}
        \toprule
        Observable                                &     Constraint                       &  Likelihood              & Reference/Comment\\
        \midrule
        $M_W$                                     &  $80.385 \pm 0.021$                  &  gaussian                & \cite{Aaltonen:2013iut} \\
        $a_\mu^{\rm exp}-a_\mu^{\rm SM}$          &  $(26.1 \pm 8.0) \times 10^{-10}$    &  gaussian                & \cite{Hagiwara:2011af,Gnendiger:2013pva} \\
        BR$(B_s \rightarrow \mu\mu)$              &  $2.9^{+1.1}_{-1.0} \times 10^{-9}$  &  from experiment         & \cite{Aaij:2013aka,Aaij:2013aka_sup} \\
        BR$(b \rightarrow s \gamma)$              &  $(3.55 \pm 0.33) \times 10^{-4}$    &  gaussian                & \cite{Amhis:2012bh} \\
        R$(B \rightarrow \tau \nu)$               &  $1.63 \pm 0.54$                     &  gaussian                & \cite{Amhis:2012bh} \\
        $m_h$                                     &  $125.0 \pm 2.0$                     &  gaussian                & \cite{CMS:2014ega} \\  %
        $m_{\widetilde{\chi}^\pm_1}$              &  $ > 45$                             &  lower limit, hard cut   & \cite{Decamp:1991uy} \\
        $m_{\tilde\chi_1^\pm}-m_{\tilde\chi_1^0}$ &  $ < 1.0$                            &  upper limit, hard cut   & see text\\
        \bottomrule
      \end{tabular}
    \end{small}
    \caption{List of the constraints used in the full likelihood for the scans. All masses are given in GeV. Experimental and theoretical errors have been added in quadrature.} \label{tab:Constraints}
  \end{center}
\end{table}

No dark matter constraints have been applied; with R-parity violation
the most natural dark matter candidates would be gravitinos or axions,
thus the standard WIMP relic abundance, direct and indirect detection
constraints do not apply. Nor have we applied constraints from direct
LHC searches for coloured sparticles. These can be avoided by pushing all the squark masses and the gluino mass up, albeit at the price of a loss of naturalness. Even the most restrictive of these bounds (for squark and gluino masses) affect the chargino--neutralino mass difference only through
small loop corrections. Direct limits on chargino--neutralino production depend intimately on the RPV coupling in question, and will be discussed in Section~\ref{sec:imp}.

As a check, scans have also been performed with a modified version of the public code {\tt SuperBayeS 1.5.1} \cite{deAustri:2006pe,Trotta:2008bp}, where the relevant quantities are calculated by {\tt SoftSusy}\footnote{Due to problems with loop contributions to the neutralino and chargino masses in {\tt SoftSusy 2.0.18}, which is the version included in {\tt SuperBayeS}, we have updated its {\tt SoftSusy} version to {\tt 3.3.7}.} and {\tt DarkSusy 5.0}~\cite{darksusy}. The conclusions from these scans agree very well with the ones reached with the setup described above.

In order to focus the scan on light charginos,
we also demand a chargino or neutralino LSP and impose an upper limit on
the chargino--neutralino mass difference. These constraints on the likelihood, which are not
from observables, are in effect restricting us to a subset of the MSSM parameter space, a
 model where the lightest chargino and neutralino have a small mass difference and one of them is the LSP. Despite being the result of a somewhat convoluted definition, there is in principle nothing that sets this model apart at the electroweak scale from other constrained models based on the MSSM field content, {\it e.g.} mSUGRA; the constraints are only different.

\subsection{Results of scan}

\begin{figure}[h!]
 \begin{center}
  \includegraphics[width=0.49\textwidth]{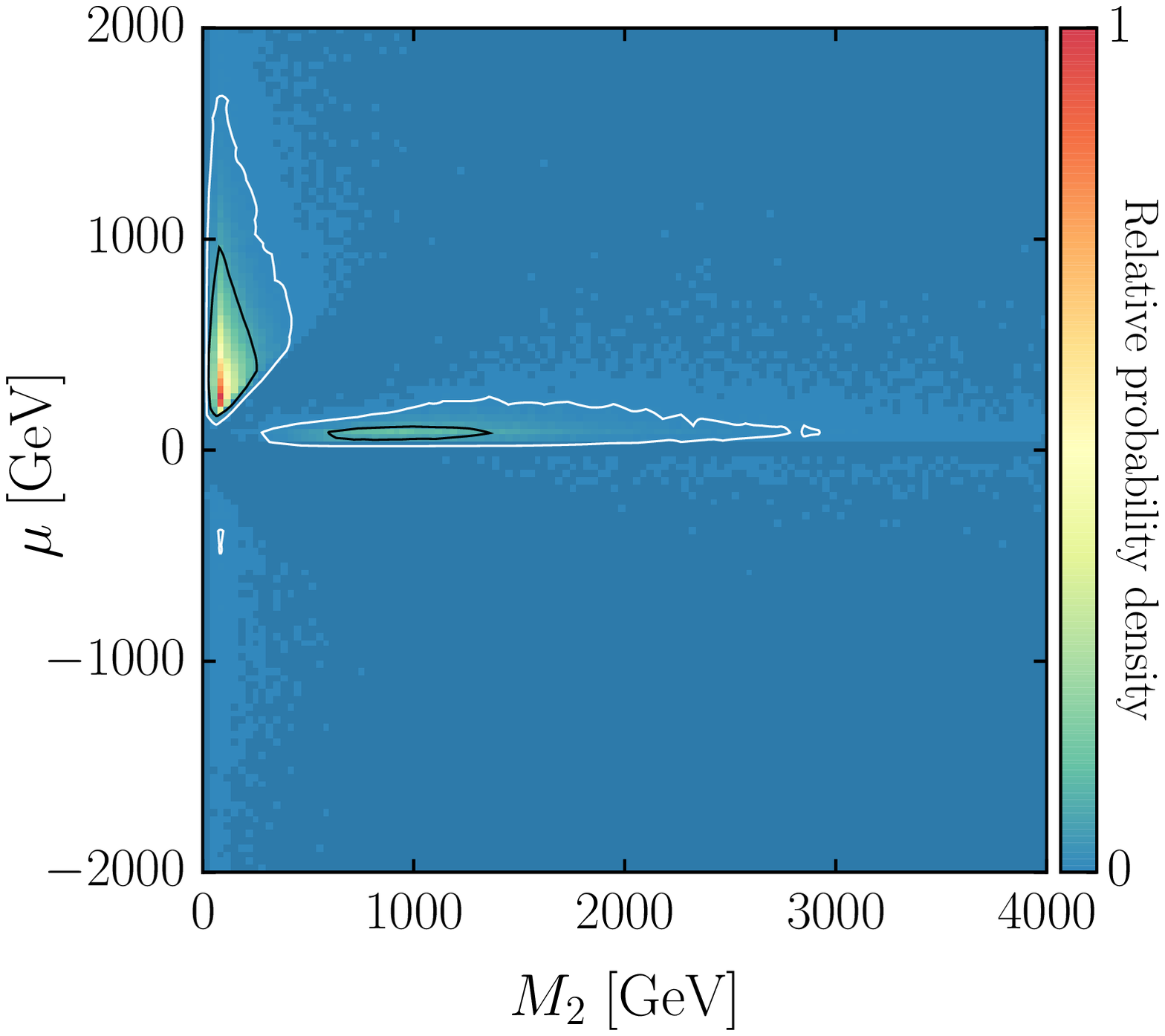}
  \includegraphics[width=0.49\textwidth]{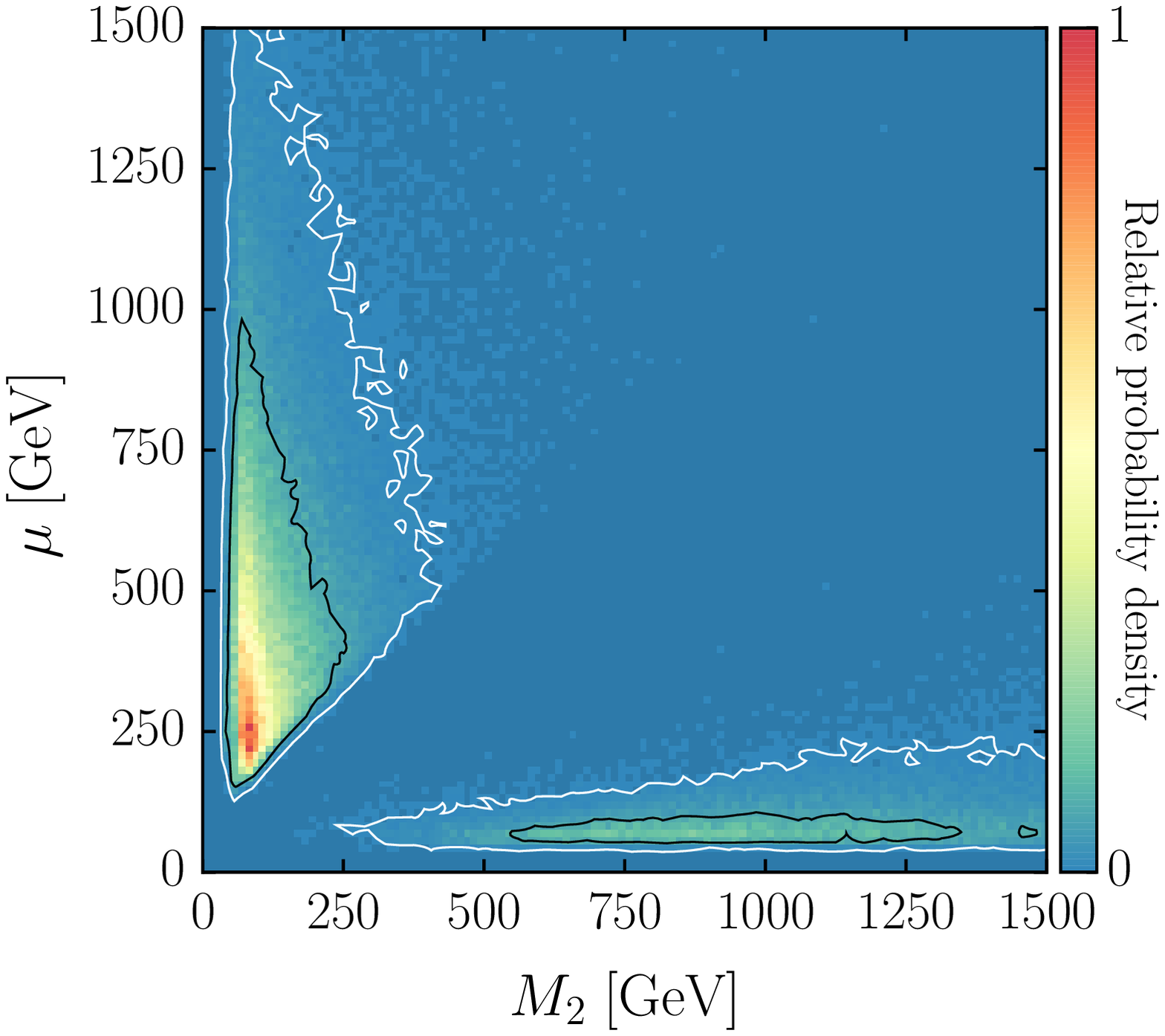}
 \end{center}
\caption{Marginalized posterior in the $M_2-\mu$ plane. The 68\%  and 95\% C.R.\ contours are shown in black and white, respectively.}\label{fig:m2mu}
\end{figure}

In Fig.~\ref{fig:m2mu} we show the marginalised posterior probability distribution in the
$M_2-\mu$ plane (left), with a higher resolution plot for smaller parameter values (right). The colour scale represents the magnitude of the probability distribution relative to its maximum point and contours of the 68\% and 95\% credible regions (C.R.) are shown in black and white, respectively. We see in principle four distinct areas, two each with wino and higgsino LSP but with different sign of $\mu$. However, the $g-2$ requirement for the muon results in a preference for a positive value of $\mu$ (same sign as $M_2$), and the resulting area of parameter space with negative $\mu$ is very small and outside the  68\% and 95\% C.R.\ contours, except for a tiny area with a wino LSP (small $M_2$).

From the scan it is clear that a wino LSP is preferred in the MSSM, $M_2<|\mu|$, when we restrict ourselves to models with small $\Delta m$. This is a consequence of the general difficulty in achieving a small mass difference from Eq.~(\ref{eq:dmhiggsino}) for the higgsinos at tree level, made worse by the relatively high Higgs mass that favours large $\tan\beta$. In addition, adding the constraint on the anomalous magnetic moment of the muon, further favours small $M_2$. We pointed out a very similar situation in the MSSM restricted to Natural SUSY models in~\cite{Bomark:2013nya}.

\begin{figure}[h!]
\begin{center}
  \includegraphics[width=0.49\textwidth]{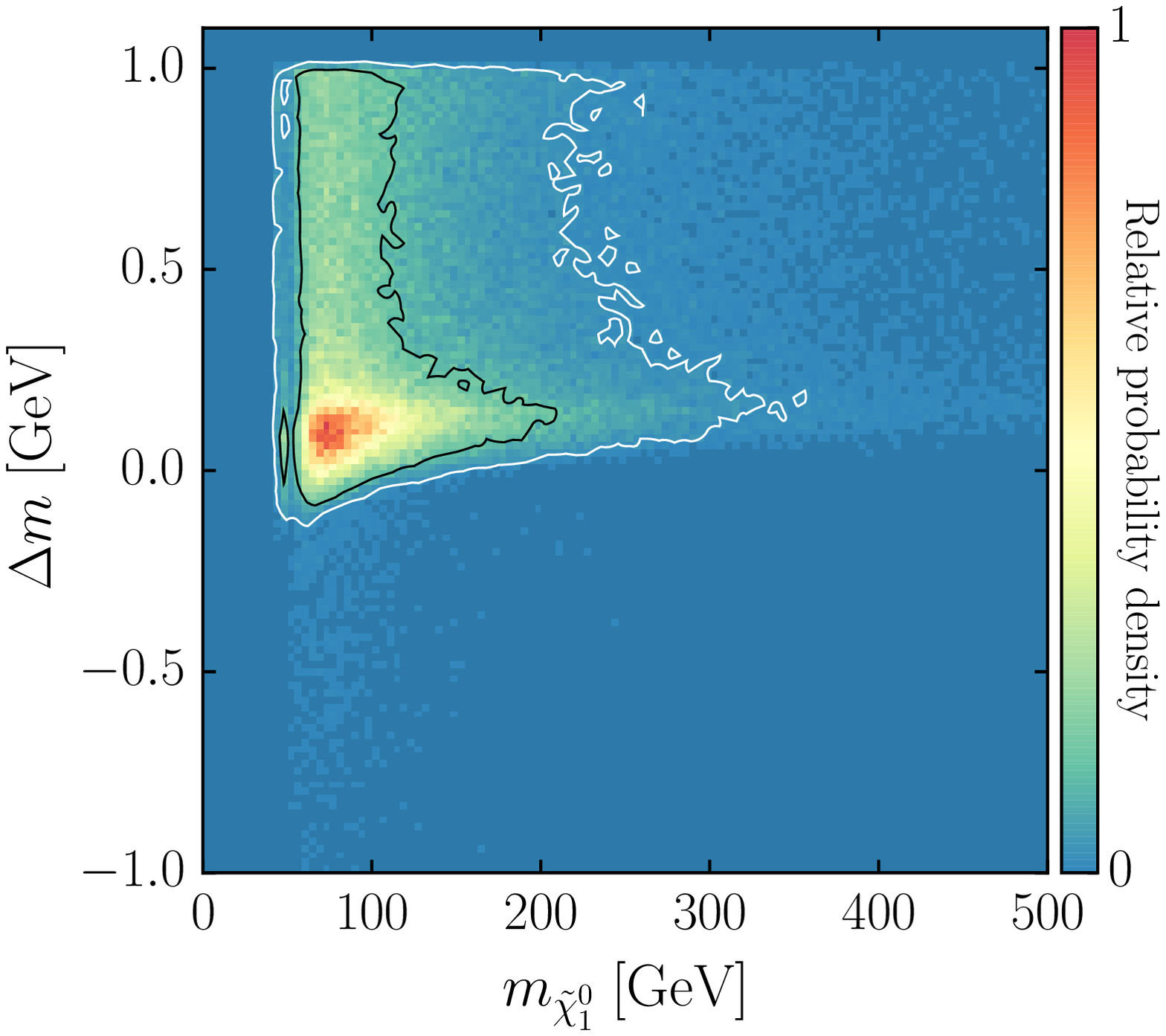}
  \includegraphics[width=0.49\textwidth]{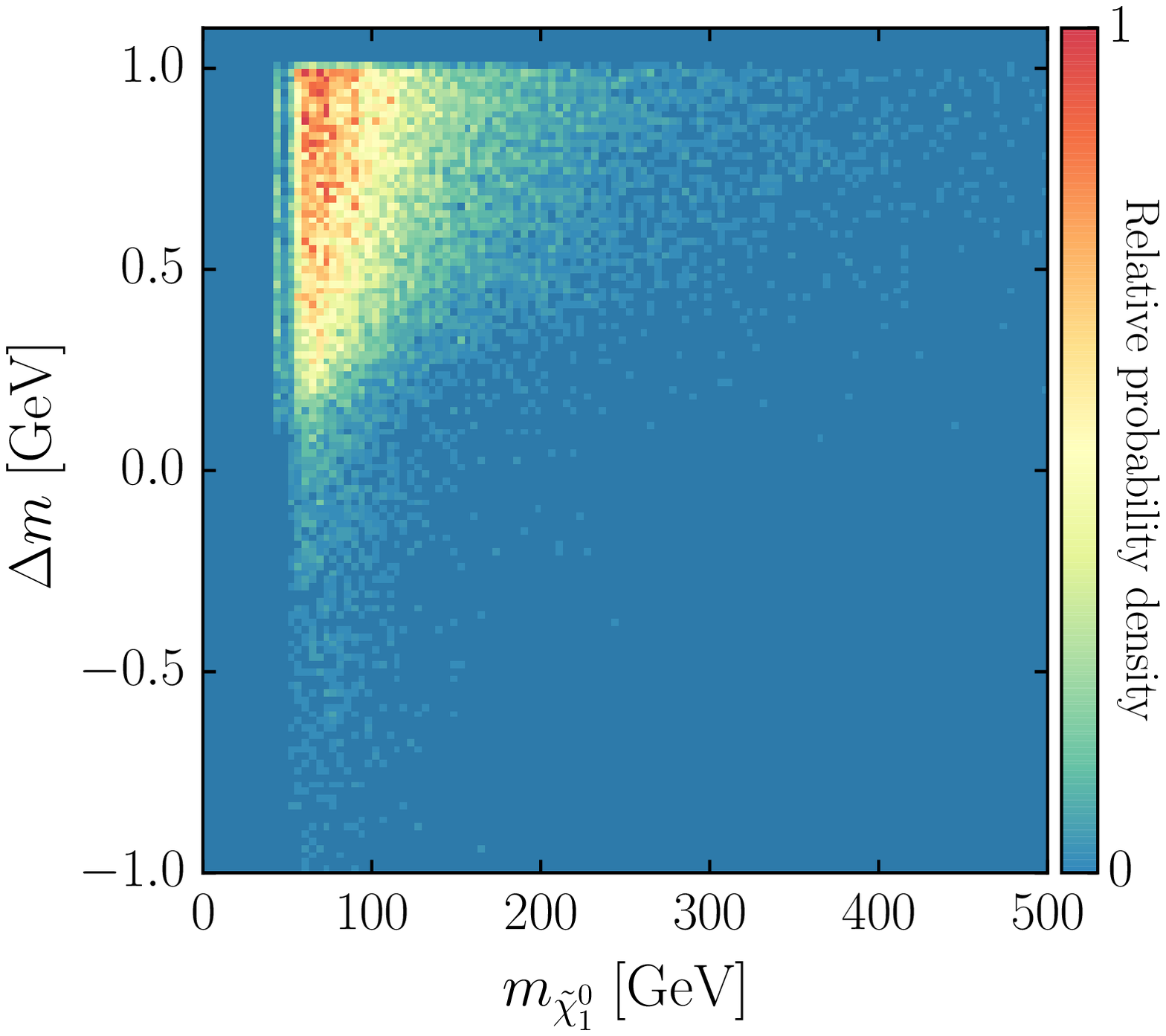}
\end{center}
\caption{Marginalized posterior in $m_{\tilde\chi_1^0}$ versus $\Delta m$ for all neutralinos (left) and for higgsino-like neutralinos only (right). The 68\% and 95\% C.R.\ contours are shown in black and white, respectively.}\label{fig:NewLimits}
\end{figure}

In Fig.~\ref{fig:NewLimits} (left) we show the marginalised posterior distribution in the $m_{\tilde\chi_1^0}-\Delta m$ plane. We see that the points naturally accumulate around the 150 MeV
mass difference given by the wino radiative correction of order $\alpha_2M_2/4\pi$. However, there is still a significant part of the preferred parameter space that has negative mass difference. To compare the wino and higgsino cases, in Fig.~\ref{fig:NewLimits} (right), we plot only the posterior points with a higgsino LSP.

The preference for a wino scenario, $M_2<|\mu|$, is even stronger in the scan using flat priors. Following the shift in priors, the posterior distributions for the mass parameters are weighted towards higher absolute values. This increases the importance of the wino radiative correction in Eq.~(\ref{eq:dmwino1loop}), thus strengthening the preference for $\Delta m$ values around 150 MeV. 
Also, the range of preferred chargino and neutralino masses is widened, with the 68\% and 95\% C.R.~in the $m_{\tilde\chi_1^0}-\Delta m$ plane extending up to $m_{\tilde\chi_1^0} \sim 650$ GeV and $m_{\tilde\chi_1^0} \sim 1100$ GeV, respectively, for $\Delta m \sim 165$ MeV.

\section{Implications for collider searches}
\label{sec:imp}

The propensity for a mass difference $\Delta m \sim m_{\pi^\pm}$ in the degenerate scenario means that in R-parity conserving models the relevant decay modes of the chargino are $\tilde\chi_1^\pm\to\tilde\chi_1^0(e^\pm\nu,\mu^\pm\nu)$ and $\tilde\chi_1^\pm\to\tilde\chi_1^0\pi^\pm$, where the latter is dominant~\cite{Chen:1996ap}. If R-parity is violated  we must also consider the  three-body decays of the chargino to three fermions via a virtual sfermion. Depending on the size and flavour of the RPV couplings, and to some extent the sfermion masses, as well as $\Delta m$, this might instead be the dominant decay channel.

Here we discuss the implication of RPV chargino decays
for collider searches, starting by describing the parameter space
where these are dominant and the consequences of current bounds on the
RPV couplings. Then, we continue with a discussion of the possibility
of displaced vertices and the limits that can be set from the absence
of massive metastable particles at the LHC. Finally, we turn to the possibility of direct searches for chargino resonances at the 13 TeV LHC.

\subsection{RPV chargino decays and current bounds}
\label{sec:current}

If the mass difference between the chargino and the neutralino is small
but still larger than $m_{\pi^\pm}$, the most important R-parity
conserving  decay channel for the chargino is
$\tilde\chi^\pm_1\rightarrow \tilde\chi^0_1\pi^\pm$. The decay width
of this channel has been given in~\cite{Chen:1996ap}. The competing
R-parity violating decay widths were given in~\cite{Dreiner:1996dd,
  Richardson:2000nt}.

\begin{figure}[h!]
 \begin{center}
\includegraphics[width=0.49\textwidth]{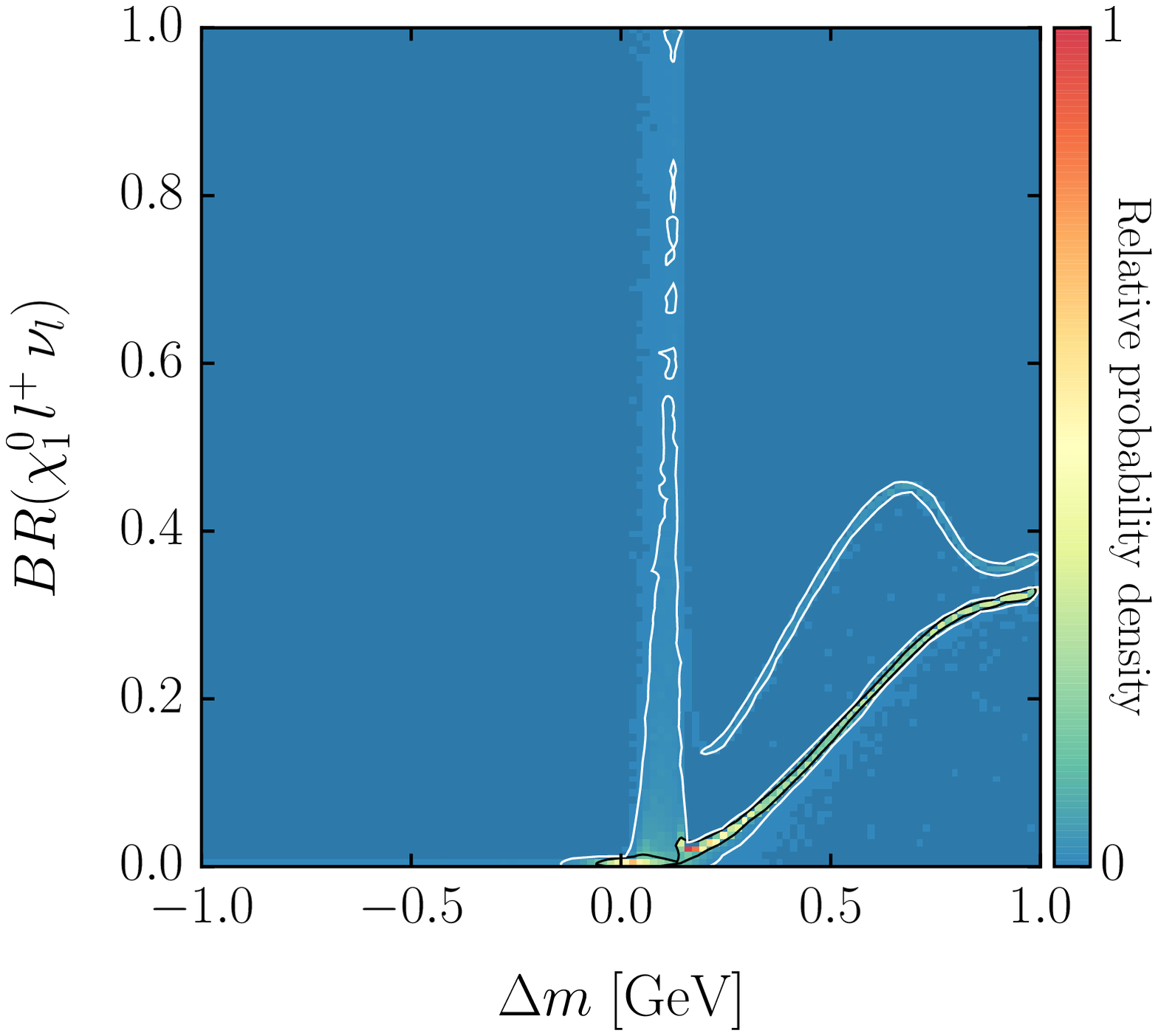}
\includegraphics[width=0.49\textwidth]{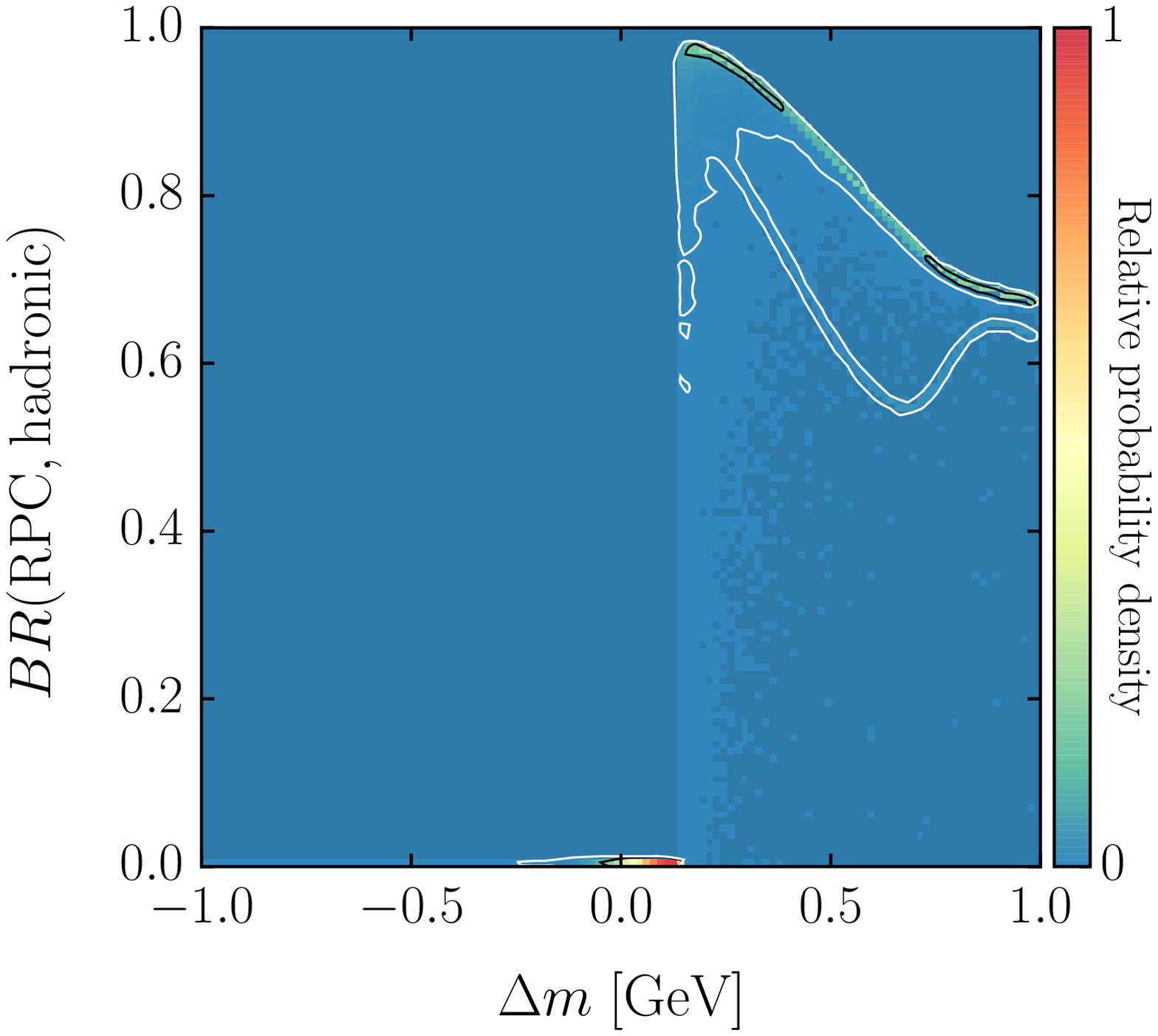}
\includegraphics[width=0.49\textwidth]{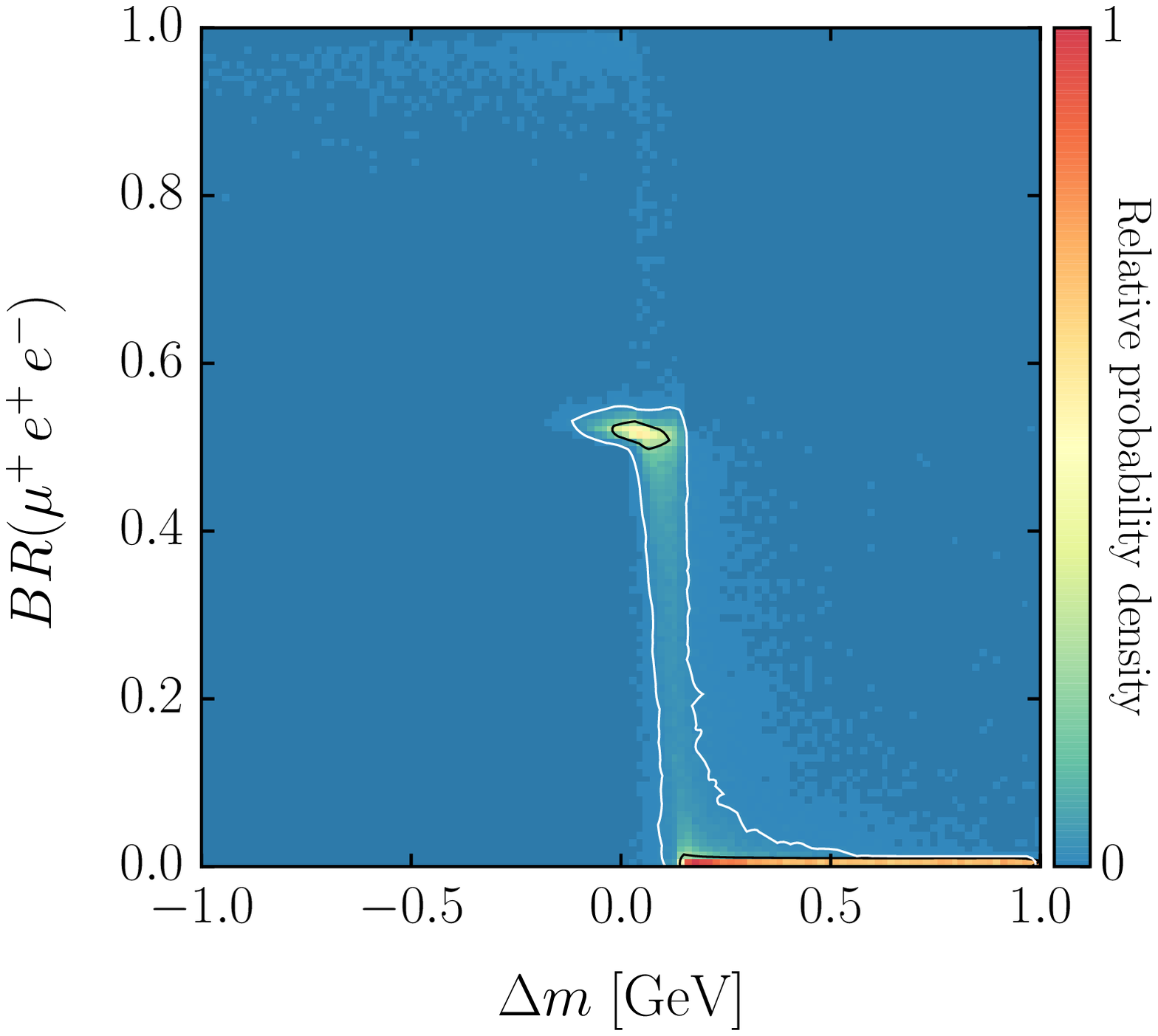}
\includegraphics[width=0.49\textwidth]{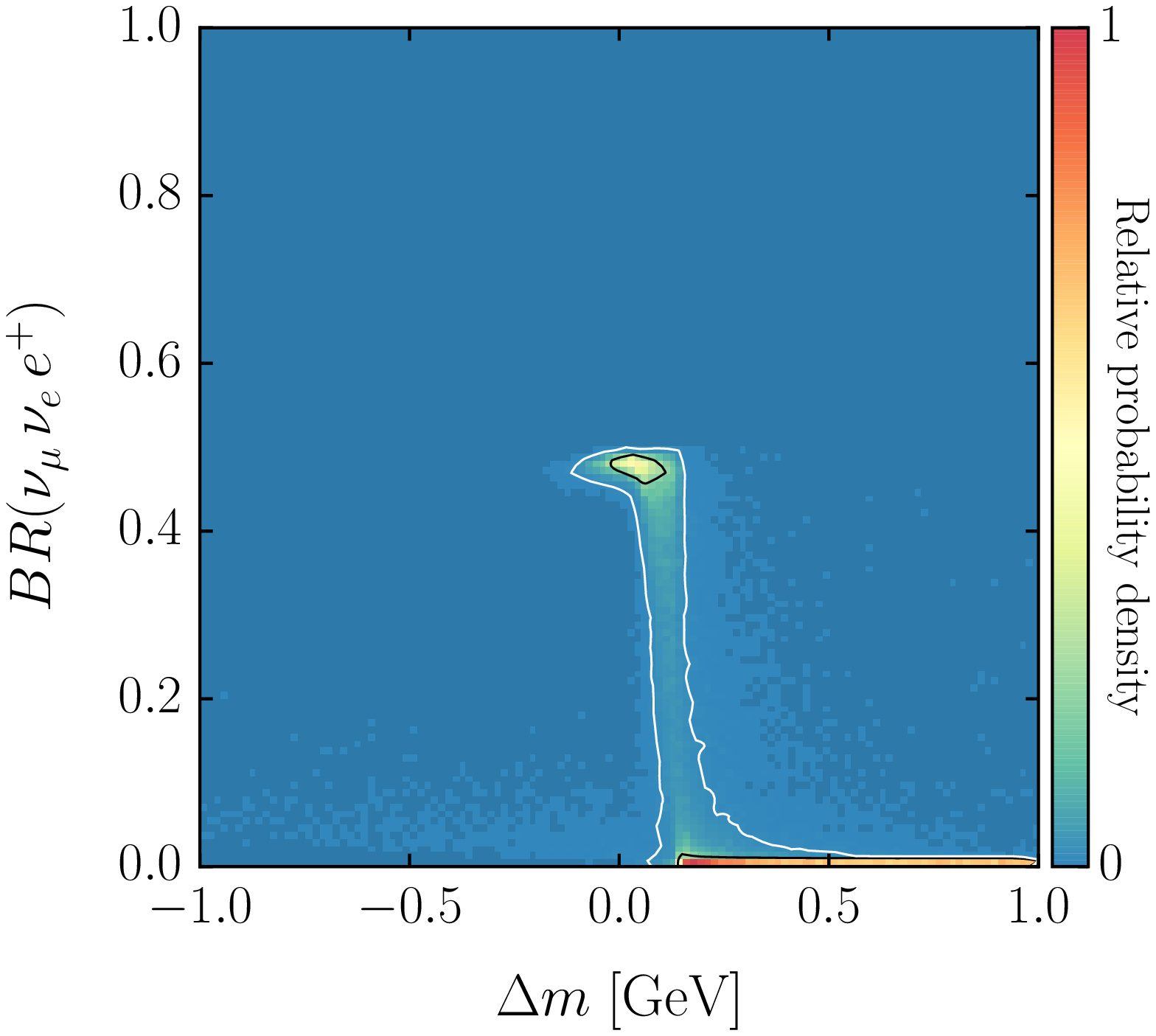}
\end{center}
\caption{Posterior distribution of $\Delta m$ versus the branching ratio for various chargino decay channels. The 68\% and 95\% C.R.\ contours are shown in black and white, respectively.}\label{fig:charginoBR}
\end{figure}

We begin by studying the effect of the $LL\bar
E$ operators on our set of posterior samples. Figure~\ref{fig:charginoBR} shows the resulting posterior distribution in the planes of relevant chargino branching ratios versus the mass difference $\Delta m$.
Here we have assumed a
dominant RPV coupling of $\lambda_{121}$. The current best
experimental limit on this coupling, from charged-current
universality, is at the weak scale $\lambda_{121} < 0.049 \times
\frac{m_{\tilde e_R}}{100\,{\rm
    GeV}}$~\cite{Barger:1989rk,Allanach:1999ic}. In the discussion
below we take the upper bounds for all couplings from~\cite{Allanach:1999ic}.  For each posterior point we choose the largest allowed value of the coupling. We remind the reader that the effect of changing the size of the coupling is a simple scaling of the RPV widths as $\lambda^2$.

The top panels of Fig.~\ref{fig:charginoBR} show the RPC decays to leptons (left) and hadrons (right). The latter dominate down to mass differences of $\sim 0.15$\,GeV, near the pion threshold. The two-pronged structure of these plots in the 95\% C.R.\ contour shows the difference between a wino and higgsino LSP, where the higgsinos prefer leptonic decays. We note that, given the discovery of a long-lived chargino with a displaced vertex in the detector, the nature of the decay products could potentially be used to discriminate between wino and higgsino.

The RPV decays shown in the lower panels of Fig.~\ref{fig:charginoBR} demonstrate that below the pion threshold there is a roughly equal splitting between the decay to three charged leptons, $e^+\mu^+e^-$, and the decay to two neutrinos $\nu_e\nu_\mu e^+$, with the three charged leptons slightly dominant. The other allowed modes, to $\bar\nu_e \mu^+\nu_e$ and $ e^+\bar\nu_{\mu}\nu_e$, both require the propagator to come from the $\bar{E}$ field which, due to its right chirality, does not couple to winos, while the coupling to higgsinos is suppressed by the small electron mass. This equitable distribution ends when the chargino becomes sufficiently light, at which point the decay to three charged leptons alone is completely dominant. This is a consequence of these charginos being higgsinos, and therefore the decay to three charged leptons is governed by the muon Yukawa coupling while the decay to two neutrinos comes from the electron Yukawa. Note though, that this scenario is outside of the 95\% C.R.\ contour of the posterior distribution.

The decay pattern we see here has important phenomenological consequences since it is relatively easy to reconstruct the chargino from three charged leptons.

Our initial choice of value for $\lambda$ maximized the RPV effect,
however, lowering $\lambda$ will affect branching ratios only in the
region between the pion mass threshold and $\Delta m = 0$ where RPC
and RPV processes compete. Here we find, fixing the sfermion mass and
changing the RPV coupling, that the total RPV decays on
average\footnote{Marginalized over the posterior sample.} reach a
branching ratio of $\sim 0.5$, when $\lambda_{121}\sim 0.05$, compared
to $\sim 1$ when $\lambda_{121}\sim 1$. For much lower values of
$\lambda_{121}$ the RPC decay is completely dominant. The scaling with
the sfermion mass, from the sfermion propagator in the RPV decays, is
$m_{\tilde f}^{-4}$. Thus an increase of the sfermion mass of a factor
4.5 has the same effect as a reduction in the coupling from 1 to 0.05.
Note also that the RPV width goes as $m_{\tilde\chi_1^\pm}^5$ while the RPC decay only depends on $\Delta m$, so for higher chargino masses the RPV decay will tend to be more dominant.

When flat priors are used, the posterior probability for having sizeable branching ratios for RPV decays increases slightly. This is mainly due to the increased probability for small $\Delta m$ values resulting from stronger wino dominance.
Also, the above-mentioned dependence of the RPV widths on $m_{\tilde\chi_1^\pm}$ and $m_{\tilde f}$ becomes more important as the range of probable values for these masses widens when using flat priors. The net result of this is a preference for RPV widths that are typically larger by a factor of a few compared to the scan with log priors.

Very similar results are found for the remaining RPV couplings.
Changing the fermion masses changes the higgsino coupling.
However, using for example $\lambda_{323}$ (the operator
with the heaviest leptons
allowed by SU(2) invariance)
we find only negligible differences
with respect to the branching ratio distributions.

Turning to the $LQ\bar D$ and $\bar U \bar D \bar D$ operators we find
that  $LQ\bar D$ sees the same behaviour as $LL\bar E$, with an
equipartition of RPV branching ratios below the pion threshold between
the final states $\ell_i^+ d_j \bar d_k$ and $\nu_i u_j \bar d_k$. The
exception to this is $L_iQ_3 \bar D_k$ where the phase space
suppression of the final state top quark implies that $\ell_i^+ b \bar d_k$ is dominant; the remaining parameter space where the chargino is heavy enough to easily decay to an on-shell top quark is small, see Fig.~\ref{fig:NewLimits}.

For $\bar U \bar D \bar D$ the generally heavy squark propagators, needed for the high Higgs mass, suppresses the RPV decays resulting in dominant RPC decays down to mass differences of $\Delta m \simeq 0.01$\,GeV even for $\lambda''=1$, and results in much longer chargino lifetimes when $\Delta m\in [10~{\rm MeV},m_\pi]$. More on this below.

\subsection{Displaced vertices from chargino decays}
\label{sec:vertex}

In the region $\Delta m < m_{\pi}$, where RPV processes can dominate, the parameters $\lambda$ and $m_{\tilde f}$  will influence the lifetime of the chargino and it is interesting to ask whether such models can give rise to detectable displaced vertices from late chargino decays.

\begin{figure}[h!]
 \begin{center}
  \includegraphics[width=0.49\textwidth]{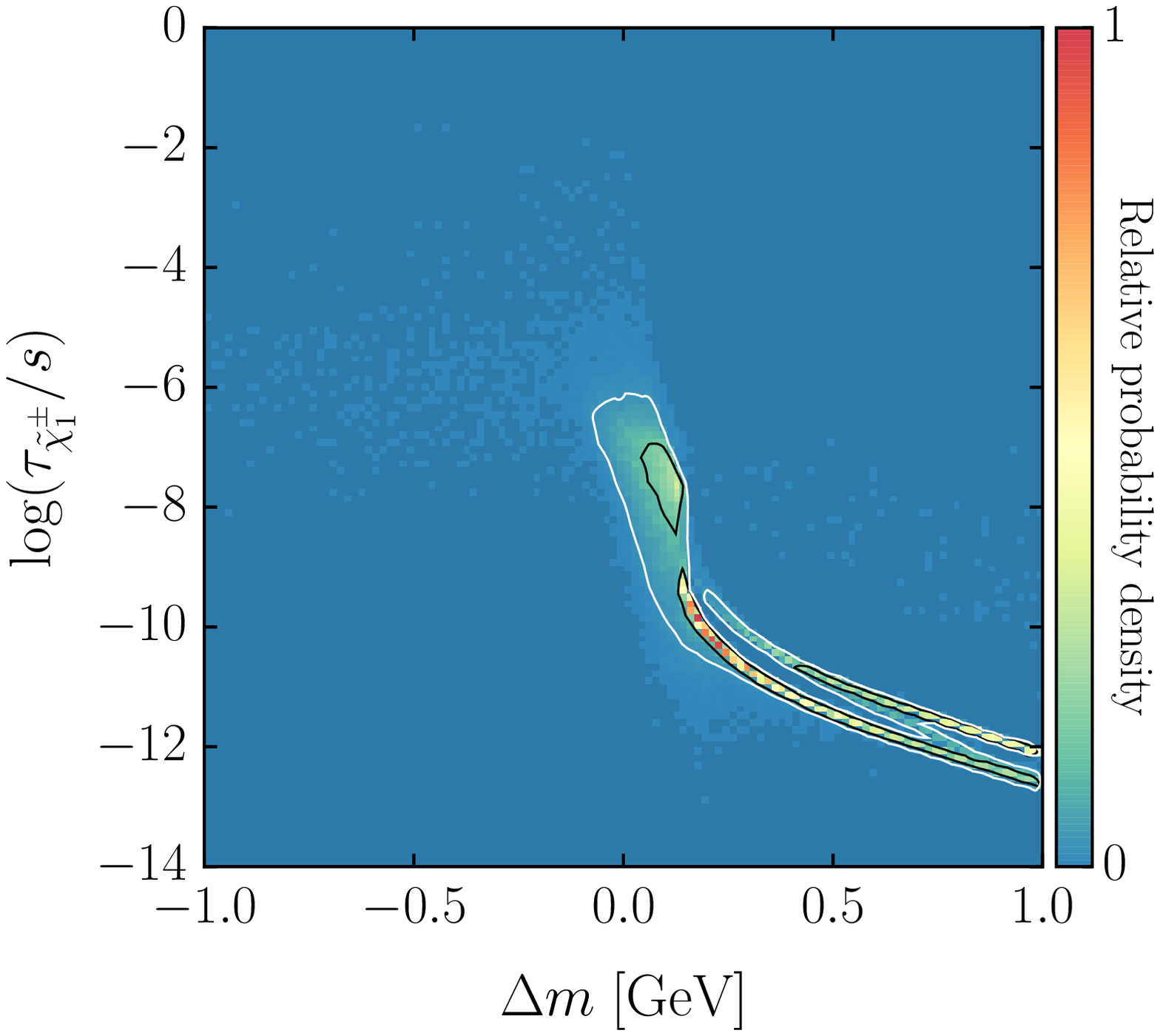}
  \includegraphics[width=0.49\textwidth]{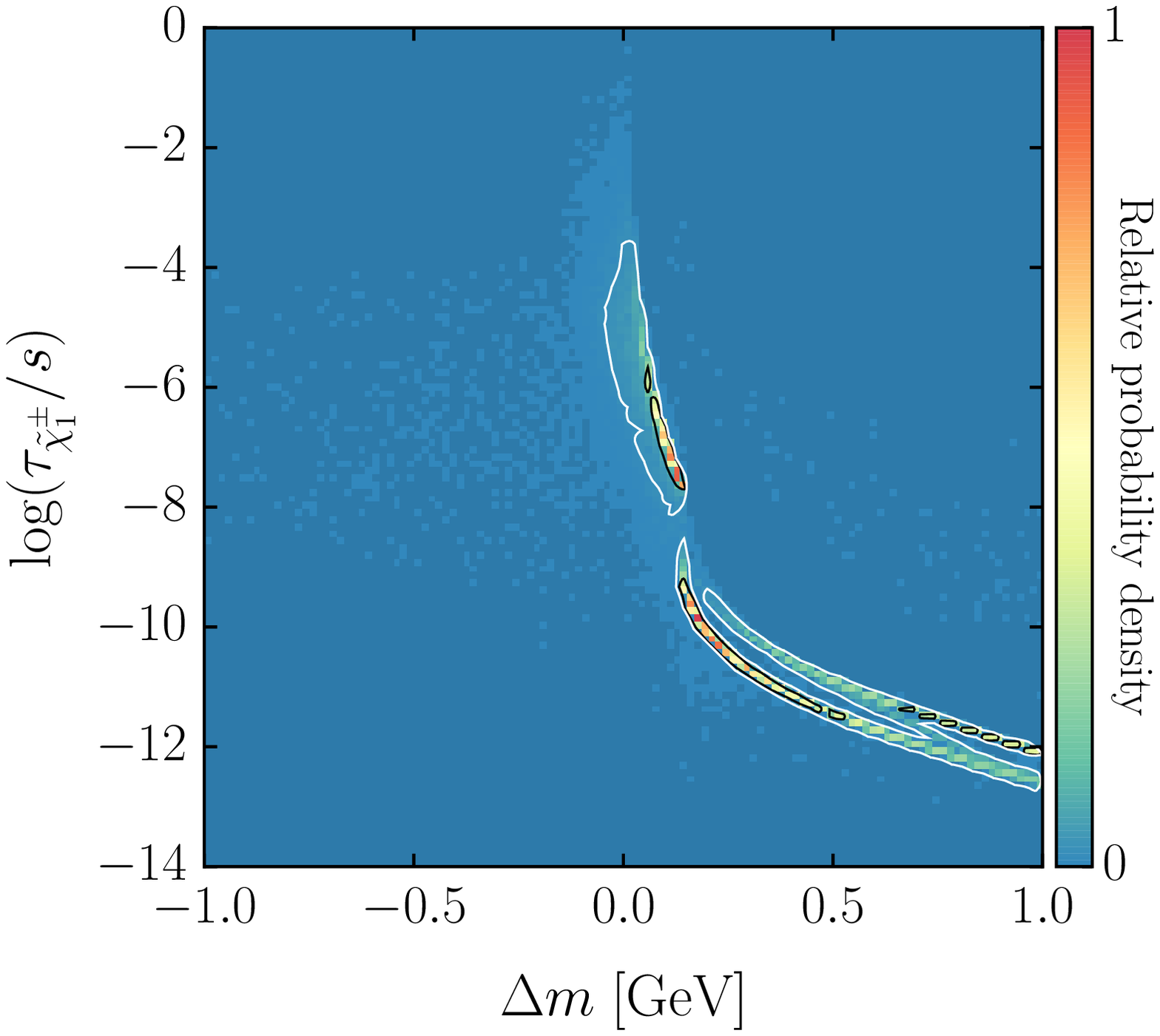}
 \end{center}
\caption{The posterior distribution of $\Delta m$ versus chargino lifetime with the $L_1L_2\bar E_1$ coupling $\lambda_{121} $ (left) and the $\bar U_3 \bar D_2 \bar D_3$ coupling $\lambda''_{323}$ (right). The 68\% and 95\% C.R.\ contours are shown in black and white, respectively.}\label{fig:Lifetime}
\end{figure}

In Fig.~\ref{fig:Lifetime} we show the posterior distribution in the plane of chargino lifetime and $\Delta m$, with the same
assuptions on the RPV coupling as above.\footnote{The limit used for
  $\lambda''_{323}$ is the perturbative limit \cite{Allanach:1999ic}
  and is not dependent on the sfermion masses.} We see that for the
$\lambda_{121}$ coupling and low mass differences (left figure)
lifetimes of $10^{-11}-10^{-7}$\,s are within the 68\% C.R.\
contours. For the $\lambda''_{323}$ coupling (right figure) the
dominance of RPC decays mentioned above drives the lifetime up even
further. The visible break is the change in RPC decays from pions to leptons at $\Delta m = m_\pi$.
Compared to the results in Fig.~\ref{fig:Lifetime}, the scan with flat priors prefers lifetimes shorter by a factor of a few in the region $\Delta m \lsim 200$ MeV, corresponding to the previously noted shift towards larger RPV widths.

Values of $c\tau>1$\,cm or $\tau> 3.3\times 10^{-11}$\,s, will give rise to a substantial number of kinked tracks in the inner detector of for example ATLAS~\cite{Barr:2002ex}, and should be detectable in the LHC experiments if a sufficient number of charginos are produced. Recent searches in ATLAS have set limits down to lifetimes of 0.06 ns ($c\tau = 1.8$\,cm)~\cite{ATLAS-CONF-2013-069} in AMSB models where the chargino is dominantly wino. We have checked our posterior sample against these limits, assuming a dominant $\lambda_{121}$ coupling, and find that in the conservative interpretation, where we assume that the charginos have a pure wino production cross section, the competitive region $\Delta m \in[0,m_\pi]$ is moderated, excluding parts of the parameter space with lifetime longer than $0.1$\,ns and chargino masses below 150\,GeV.

Nevertheless, significant parts of the parameter space remain within the 68\% C.R.\ with a 50\% branching ratio to three charged leptons. This surviving region prefers somewhat heavier charginos than the posterior sample following the scan, relatively high $\tan\beta$, and sees lifetimes in the region $0.1-0.01$\,ns, which can be within the future reach of the LHC experiments looking for displaced vertices. We study three such points, {\tt RPV\_C1}, {\tt RPV\_C2} and {\tt RPV\_C3}, as benchmark points in the next section. Details of these points are given in Table~\ref{tab:benchmark}.

\begin{table}[h!]
  \begin{center}
    \begin{small}
      \begin{tabular}{lccc}
        \toprule
        Point  & {\tt RPV\_C1} & {\tt RPV\_C2} & {\tt RPV\_C3}  \\
        \midrule
        $m_{\tilde\chi_1^\pm}$ & 252.1 & 327.7     & 526.4 \\
        $\Delta m$                    & 0.119 & 0.108     & 0.182 \\
        Wino                             & 0.990 & 0.986     & 0.989 \\
        Higgsino                       & 0.142 & 0.166     & 0.148 \\
	\midrule
        $M_1$                           & 944.1  & -1082.0 & -728.4\\
        $M_2$                           & 235.4  & 311.4     &  502.3\\
        $M_3$                           & 1627.6 & 560.6    & 3418.6\\
        $\mu$                            & 668.0   & 668.5    &  913.2 \\
        $m_{A^0}$                     &3430.3  & 2775.5  & 3220.5\\
        $m_{\tilde l}$                  & 503.5  & 434.6     &  757.6\\
        $m_{\tilde q}$                 & 2156.2 & 2517.0   & 4742.9\\
        $m_{\tilde q_3}$             & 6429.4 & 4951.8   & 1424.6\\
        $A_0$                             & -25.8   & 2775.5    & 1498.1\\
        $\tan\beta$                     & 47.1    & 55.4         & 46.2\\
        \bottomrule
      \end{tabular}
    \end{small}
    \caption{Summary of the properties of the benchmark points studied. All masses are in GeV.}  \label{tab:benchmark}
  \end{center}
\end{table}

The dependence of these conclusions on which RPV coupling is dominant,
for a particular class of operators, say $LL\bar E$, is very weak
because of the assumptions in the scan, namely that of common
weak-scale sfermion soft masses. For the $LQ\bar D$ operators the
situation is similar to $LL\bar E$, with somewhat shorter
lifetimes predicted. This is because the dominant RPV decay channels, $\ell_i^+
d_j \bar d_k$ and $\nu_i u_j \bar d_k$, have the same slepton
propagators as the dominant $LL\bar E$ decays; at the same time the
upper bound on the coupling size is less restrictive, since, for most
couplings, this depends
on large squark masses. After applying the ATLAS limits from displaced
vertices, we still have significant regions of the parameter space
with dominant decays to $\ell_i^+ d_j \bar d_k$ and $\nu_i u_j \bar
d_k$ within the 68\% confidence region. One exception here is
$\lambda'_{111}$ where the bound on the coupling from neutrino-less double beta decay is sufficiently strong to exclude most points with dominant RPV decays, except points that also have very large slepton masses.

For $\bar U \bar D \bar D$ the heavy squark propagators reduce the
decay width for RPV decays, leading to longer chargino lifetimes as compared to $LL\bar E$ and $LQ\bar D$. As a result, the displaced vertices search excludes all regions of the preferred parameter space, the 95\% C.R., where RPV decays from $\bar U \bar D \bar D$ are substantial, having above 1\% branching ratio.

The restrictions on the individual RPV couplings, from above by the indirect bounds taken from \cite{Allanach:1999ic}, and from below by the ATLAS lifetime bounds, when seen together, are quite severe for all the preferred regions in the parameter space discussed in Section~\ref{sec:scan}. As an example we list maximum values for the $LL\bar E$ couplings in Table~\ref{tab:maxlambda} for the benchmark points {\tt RPV\_C1}, {\tt RPV\_C2} and {\tt RPV\_C3}. The minimum values for  {\tt RPV\_C1} vary in the range 0.222--0.228. Since the ATLAS bound extends up to  chargino masses of 500 GeV, it is difficult to find points with low fine-tuning that completely escape this bound. For comparison, a benchmark point  {\tt RPV\_C2} with higher chargino mass has minimum values in the range 0.055--0.059, where the less severe bounds are in part also caused by lower slepton masses. The point {\tt RPV\_C3} has no such lower bound, since it has a chargino mass above 500 GeV and is thus outside the reach of the ATLAS search. From these results, we conclude that certain combinations of couplings and points with dominant RPV decays are unlikely in the context of natural models; in particular a dominant $\lambda_{133}$ coupling is hard to realize.

\begin{table}[h!]
  \begin{center}
    \begin{small}
      \begin{tabular}{lccccccccc}
        \toprule
        Point/Coupling    &   $\lambda_{121}$ &  $\lambda_{122}$ & $\lambda_{123}$  & $\lambda_{131}$ & $\lambda_{132}$ & $\lambda_{133}$ & $\lambda_{231}$& $\lambda_{232}$ & $\lambda_{233}$\\
        \midrule
       {\tt RPV\_C1} & 0.244 & 0.244 & 0.260 & 0.309 & 0.309 & 0.013 & 0.349 & 0.349 & 0.372\\
        \midrule
        {\tt RPV\_C2} & 0.215 & 0.215 & 0.221 & 0.272 & 0.272 & 0.011 & 0.307 & 0.308 & 0.316\\
         \midrule
        {\tt RPV\_C3} & 0.369 & 0.369 & 0.388 & 0.467 & 0.467 & 0.016 & 0.527 & 0.528 & 0.554\\
         \bottomrule
      \end{tabular}
    \end{small}
    \caption{Maximum allowed values of the $LL\bar E$ couplings for the benchmark points discussed in the text.}  \label{tab:maxlambda}
  \end{center}
\end{table}

\subsection{LHC resonance searches}
\label{sec:lhc}

To study the possibility of observing RPV chargino decays at the LHC we generate events at 13 TeV using {\tt PYTHIA 8.1.80}~\cite{Pythia8}, with {\tt FASTJET 3.0.6}~\cite{FastJet} for jet reconstruction
using the $k_t$-algorithm~\cite{ktJet1,ktJet2} with jet radius $R=0.4$.\footnote{In order to calculate decay widths and branching rations for the sparticles, including RPV operators, we use {\tt PYTHIA  6.4.25}~\cite{PYTHIA}, modified to include the
$\tilde\chi_1^\pm\to\tilde\chi_1^0\pi^\pm$ decay.}  We use the single dominant coupling
approximation.\footnote{We note that large R-violating hierarchies can be
expected in many models, similarly to the large hierarchies in the
Yukawa couplings that generate fermion masses.} As the search for RPV decays of charginos is similar to the corresponding decays of neutralinos, we refer to~\cite{Bomark:2011ye} for details on the analysis. Here we will mostly focus on the differences that appear when charginos are involved.

\subsubsection*{$LL\bar E$ operators}
The most obvious difference between neutralinos and charginos, as discussed in the previous section, is that for  $LL\bar E$ operators, the chargino has a large branching ratio for $\tilde\chi^\pm_1\to
lll,l\nu\nu$ via a virtual sneutrino or slepton, raising the possibility to observe a resonance of three charged leptons. Since for $\lambda_{ijk}$
$SU(2)$ invariance requires $i\neq j$, there are always at least two
different flavours in the final state leptons.  Thus, the pure
leptonic combinations that are most relevant for collider searches
are $e^+e^-\mu^+$ ($\lambda_{121}$), and $\mu^+\mu^-e^+$ ($\lambda_{122}$). The
remaining $LL\bar E$ operators have some tau flavour in them, which necessarily smears out any resonance peak.

We employ the cuts used in~\cite{Bomark:2011ye}, which require many
high $p_T$ leptons and missing energy:
\begin{itemize}
\item Three isolated leptons with $p_T>70,20,20$\,GeV within the detector's geometric acceptance.
\item Missing transverse energy $E_T^{\rm miss}>100$\,GeV.
\end{itemize}
We simulate LHC data for benchmark points {\tt RPV\_C1} and {\tt RPV\_C3} taken from our scan, with the largest possible value of the RPV couplings allowed for the slepton masses of those points. These couplings were given in Table~\ref{tab:maxlambda}.\footnote{Since for our benchmark points the RPV decay is clearly dominating, relatively small changes to $\lambda$ will not affect the result, larger changes would decrease the signal as $\lambda^2$.} A full PYTHIA simulation of neutralino and chargino pair production is performed, meaning that both the lightest neutralino and chargino will decay through the RPV couplings. The presented signal distributions  are therefore a superposition stemming from both types of decays.

\begin{figure}
 \begin{center}
  \includegraphics[width=15cm]{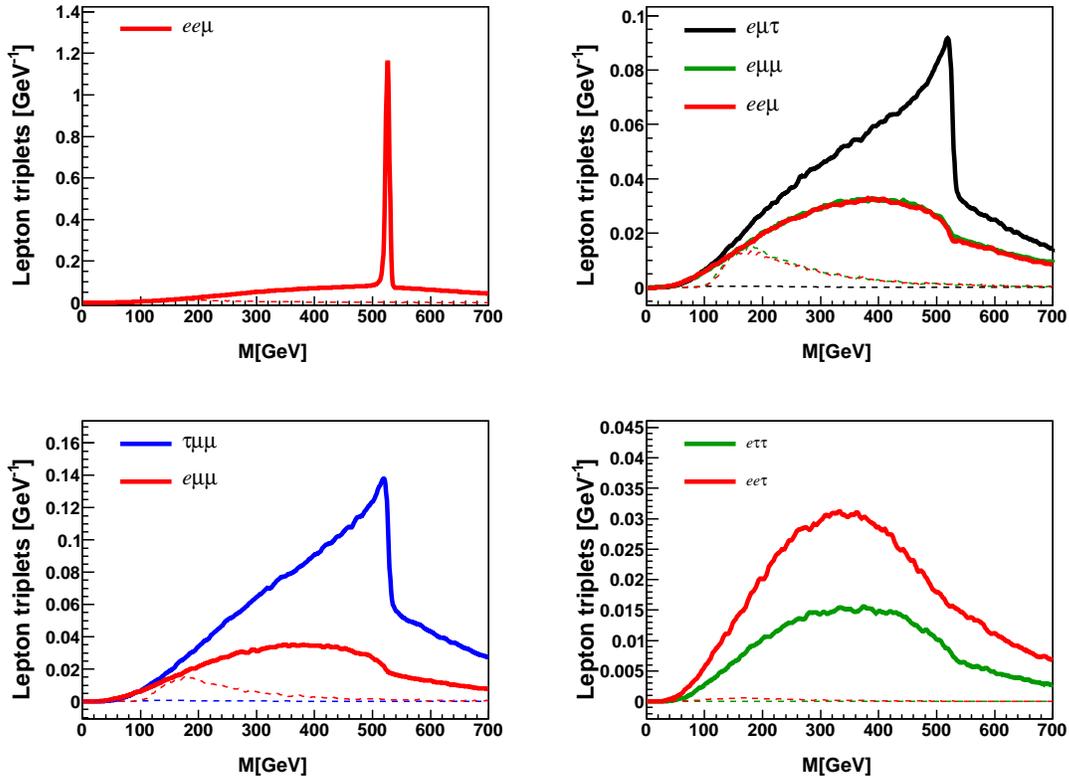}
 \end{center}
\caption{Various flavour combinations of tri-lepton invariant masses for the $L_1L_2\bar E_1$, $L_1L_2\bar E_3$, $L_2L_3\bar E_2$ and $L_1L_3\bar E_3$ couplings. The thin dashed lines give the dominant background and the distributions are normalized to 1 fb$^{-1}$ of integrated luminosity.}\label{fig:TriLep}
\end{figure}

In Fig.~\ref{fig:TriLep} we show the resulting tri-lepton invariant
mass distributions for the benchmark point {\tt RPV\_C3} with the $L_1L_2\bar E_1$, $L_1L_2\bar E_3$, $L_2L_3\bar E_2$ and $L_1L_3\bar E_3$ operators (from top left to bottom right) for signal and dominant background,\footnote{For the tri-lepton distributions we use WZ production for the background (see below for a discussion of the backgrounds) while for the distributions including tau-jets we use ZZ production since WZ will not give tau-jets in addition to the three leptons required to pass the cuts.} normalized to 1 fb$^{-1}$ of integrated luminosity. The $\tau$ in these plots refers to hadronic jets stemming from
taus. As expected, $L_1L_2\bar E_1$ shows a clearly identifiable
peak at the chargino mass of 526 GeV above the combinatorial background. For $L_1L_2\bar E_3$ and $L_2L_3\bar E_2$ we see clearly identifiable features in the distributions with taus as well as identifiable kinks in the purely leptonic distributions. For these couplings the chargino should be observable as a resonance up to quite high masses at 13--14 TeV.
Due to the large content of taus, $L_1L_3\bar E_3$ appears more challenging; there is a small kink in the $e\tau \tau$ distribution, but one should bear in mind that these plots are produced from $10^6$ generated events (before cuts) and this is an unrealistically high number as compared to experimental expectations, so it remains an open question whether these smaller features can actually be observed.

From preliminary studies by ATLAS of the tri-lepton reach in supersymmetry models at 14 TeV~\cite{ATLAS13TeVTriLepton}, the main background to the tri-lepton resonance is expected to be di-boson production, in the absence of a $Z$-boson veto, and tri-boson and $t\bar t$ with said veto. Comparing the total NLO cross section of 175 pb for di-boson production at 13 TeV~\cite{Campbell:2011bn}, to the total neutralino and chargino pair-production cross section for {\tt RPV\_C1} and {\tt RPV\_C3} of 0.99 pb and 49.9 fb, respectively, calculated using {\tt Prospino 2.1}~\cite{Beenakker:1999xh}, both benchmark points seem promising for early discovery.\footnote{We note here that some of the four-lepton searches in \cite{Aad:2014iza} may be sensitive to chargino masses as low as in {\tt RPV\_C1}, however, it is difficult to fully understand their impact without a detailed simulation of each search, which is outside the scope of the present paper. {\tt RPV\_C3} should be safely outside of their reach. }


In order to quantify this further, we can use the selection efficiency
of the di-boson background and signal for the $ee\mu$ final state
under the cuts given above,\footnote{In order to explore the many
  possible RPV couplings we have kept our cuts very
  generic. Improvements for specific couplings can certainly be
  made. However, we feel a more detailed study of cuts and
  sensitivity, than our rough estimate, would be better done within
  the experiments.} which we find to be $5.7\times 10^{-5}$ for the background,
and 0.24 and 0.47 for {\tt RPV\_C1} and {\tt RPV\_C3},
respectively.\footnote{Calculated using leading order matrix elements
  in the Monte Carlo simulation.} Using an $S/\sqrt{B}$ criterion to
test sensitivity, where $S$ is the number of signal events and $B$ the
background, and requiring a minimum of five signal events, a simple
event counting experiment will need 21 pb$^{-1}$ and 0.45 fb$^{-1}$ of
data to reach discovery level for the two benchmarks.  This is clearly
an early discovery opportunity for Run II of the LHC.
Because of the very low SM background, in order to observe the chargino peak above
the combinatorics of the events with supersymmetric origin, slightly
more statistics would be needed. Results for the $e\mu\mu$ state are
very similar.

More detailed studies could be performed
by the LHC experiments, to take into account the missing energy
resolution and the lepton reconstruction efficiency, which
should be high at the transverse momenta of
interest, and to include a full
background analysis. Here we took into account only the most
significant background. These however go beyond the scope of this
paper.

Looking for the tri-lepton peak is crucial for identifying the scenario giving rise to a multi-lepton excess. It would tell us that we have a charged particle that decays to a specific combination of three leptons, thus violating lepton number, as well as the mass of that particle, and potentially the spin of the particle through angular distributions of the decay products. As already mentioned, the distributions of Fig.~\ref{fig:TriLep} may help with these identifications also for those operators that do not give rise to a clear peak.

As in the case of neutralino decays discussed in~\cite{Bomark:2011ye}, there are also interesting features in the di-lepton invariant mass distributions. The advantage with di-lepton distributions is that we can employ same-sign subtraction to practically remove the combinatorial background and hence reveal features otherwise invisible. The combinatorial background consists of lepton pairs that come from different parts of the event and are hence expected to be uncorrelated. As a result, the charges of these lepton pairs should be uncorrelated and therefore, taking for example the difference, $m_{e^+e^-}-m_{e^-e^-}-m_{e^+e^+}$ for the $ee$ invariant mass, should remove the combinatorial background.

\begin{figure}[h!]
 \begin{center}
  \includegraphics[width=15cm]{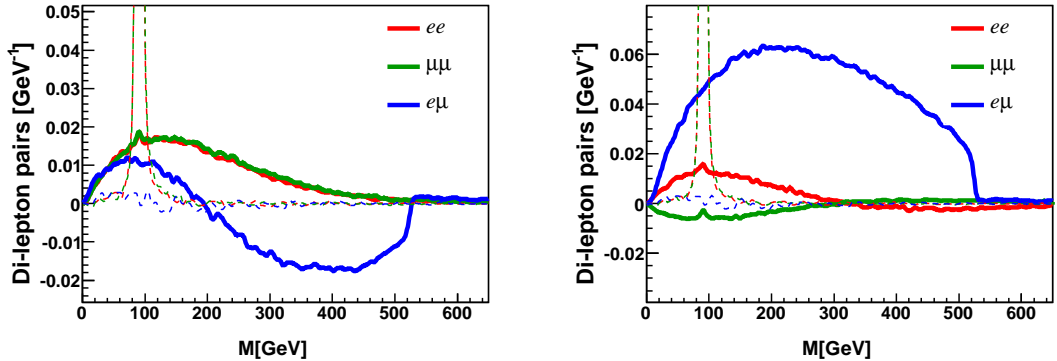}
 \end{center}
\caption{Di-lepton invariant masses for $L_1L_2\bar E_3$ (left) and $L_2L_3\bar E_1$ (right). The thin dashed lines give the dominant background and the distributions are normalized to 1 fb$^{-1}$ of integrated luminosity.}\label{fig:DiLep}
\end{figure}

As can be seen in Fig.~\ref{fig:DiLep}, where the resulting di-lepton
invariant mass distributions for $L_1L_2\bar E_3$ and $L_2L_3\bar E_1$
(signal and WZ background for 1 fb$^{-1}$ of integrated luminosity) are shown, this works as expected. For neutralino
decays, these distributions were discussed in detail
in~\cite{Bomark:2011ye}; here we focus on those features that are
relevant for decaying charginos. The most striking feature can be seen
in the left panel of Fig.~\ref{fig:DiLep} where the $e\mu$ invariant
mass distribution becomes negative. This effect is due to the
structure of the couplings; the two distinct flavours in the lepton
superfield doublets $L$ will give rise to contributions in the
same-sign invariant mass distributions from the chargino decays. After same-sign subtraction
this will show up as a negative contribution in the distribution, and
interfere with the positive contributions from both decaying charginos and neutralinos.

Following~\cite{Bomark:2011ye}, the cleanness of the di-lepton distributions can be used to extract a substantial amount of information about the scenario at hand, including the chargino/neutralino mass and the flavours of the couplings, also in the difficult cases where the decays are dominated by taus. In addition one could hope to establish the ratio of neutralino to chargino decays in the final distribution.

To illustrate this, let us look at the $ee$ and $\mu\mu$ distributions
in the right-hand panel of Fig.~\ref{fig:DiLep}. The negative perturbation in the
$\mu\mu$ channel shows that we have a decaying chargino, while the
absence of a clear cutoff as in the $e\mu$ channel indicates that one
of the muons comes from a decaying tau. This means that our coupling
must have an $L_2L_3$ component. The large peak in the $e\mu$ channel
indicates that the last component of the RPV operator must be $\bar
E_1$, and it also gives us the chargino/neutralino mass. Finally, the
positive perturbation in the $ee$ distribution is caused by neutralino and chargino
decays with an $ee$ pair where one of the electrons comes from a
decaying tau and the other from the $\bar E_1$ operator.

We can now compare the height of the perturbation in the $ee$ distribution
to the one of the negative perturbation in the $\mu\mu$ distribution. Since
the contribution from chargino decays is the same in both distributions,
the difference in height between the positive and negative perturbations, when compared to the (negative) height of the $\mu\mu$ distribution, reveals the ratio of neutralino to chargino decays in our event
sample. In this case charginos are slightly more common than neutralinos.

\subsubsection*{$LQ\bar D$ operators}
\begin{figure}[h!]
 \begin{center}
   \includegraphics[width=15cm]{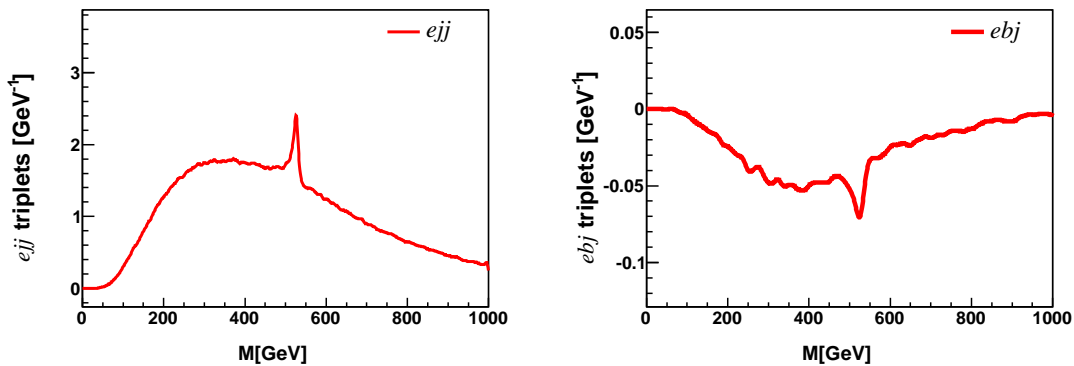}
 \end{center}
\caption{Lepton plus dijet invariant mass distributions for $L_1Q_1\bar D_1$ (left), and the same for $L_1Q_3\bar D_1$ including same-sign subtraction (right). Signal only, normalized to 1 fb$^{-1}$ of integrated luminosity.}\label{fig:ljj}
\end{figure}

In Fig.~\ref{fig:ljj} (left) we show the distribution of the invariant mass of $ljj$ combinations for the {\tt RPV\_C3} benchmark point with the $L_1Q_1\bar D_1$ operator for signal only, with arbitrary normalization. Since light quark flavour, and thus charge, is impossible to determine experimentally from jet physics, chargino resonances decaying through the $LQ\bar D$ operators have already received significant attention in searches for leptoquarks and neutralinos with RPV decays. We will therefore limit our discussion here. In terms of chargino masses the same cross section limits should apply as for neutralinos.

One exception that is worth mentioning are operators of the type $L_iQ_3\bar D_k$ that cause relatively light neutralinos to always decay to neutrino plus jets, $\nu_i b \bar d_k$, while heavier neutralinos can decay to $l^-_i t\bar d_k$. Light charginos will always have the possibility to decay to charged lepton plus jets, $l^-_i b \bar d_k$, thus improving the detectability significantly with both a lepton and a  b-tag, and potential charge identification, for the resonance reconstruction. This can be further improved by same-sign subtraction for the electron and b-jet pair in order to reduce combinatorial background. We show the resulting distribution of
\begin{equation}
m_{e^-\bar bj}+m_{e^+bj}-m_{e^+\bar bj}-m_{e^-bj},
\end{equation}
for the $L_1Q_3\bar D_1$ operator in Fig.~\ref{fig:ljj} (right). Here the chargino contribution has been subtracted, showing up as the downward fluctuation in the distribution at $\sim 530$ GeV.

We also note that models with large $\lambda'_{111}$ and $m_{\tilde\chi_1^0}$ from 500 GeV to 1 TeV were recently studied \cite{Allanach:2014lca} as a possible explanation of the deviations from the Standard Model seen in the $eejj$ channel in CMS data~\cite{Khachatryan:2014dka}, however, this requires sleptons with masses around 2 TeV, which are very disfavoured in our search due to the naturalness bias and the constraint on $(g-2)_\mu$ that we have applied. Even so, it is worth pointing out that a degenerate chargino--neutralino pair is not impossible in this scenario, and that such a chargino can be searched for as a resonance in the $l^+ d \bar d$ channel.

\subsubsection*{$\bar U\bar D\bar D$ operators}
For the $\bar U\bar D\bar D$ operators the situation is similar to the $LQ\bar D$ operators. The chargino decays as $\tilde\chi^+_1\to u_iu_jd_k,\bar d_i\bar d_j\bar d_k$. For light flavours a chargino resonance will look just like a neutralino (or gluino) resonance decaying into three quarks, $\tilde\chi_1^0\to u_i d_j d_k$, and will be equally challenging to reconstruct, although some improvement could be had from the sophisticated use of jet clustering algorithms  \cite{Butterworth:2009qa}.  Experimental bounds on three-jet resonances do exist \cite{Chatrchyan:2013gia}, but these have been interpreted only in terms of gluino production and RPV decay, where the production cross sections are much higher. As we have seen in Section~\ref{sec:vertex}, RPV decays of charginos through $\bar U\bar D\bar D$ operators should be heavily suppressed compared to RPC decays in scenarios that feature naturalness, and will thus be even more difficult to discover.

We end by pointing out that for the operators $\bar U_3 \bar D_j \bar D_k$, a neutralino decay requires a top quark in the final state, while the chargino can also decay to three down quarks, of which at least one is a bottom quark. As discussed above in Section~\ref{sec:vertex} it seems difficult to realise a scenario where the chargino is much lighter than 500 GeV and decays dominantly through RPV couplings. Thus, this inherent difference may have only marginal applicability since the decay to top quarks will not be very kinematically suppressed at such high masses. The final states unique to the chargino are the decays to two b-quarks and one light quark, with the experimental signature $bbj$, and the decays to two top quarks and a light down quark, $ttj$, which are possible for the $\bar U_3 \bar D_1 \bar D_3$ and $\bar U_3 \bar D_2 \bar D_3$ operators.

\section{Conclusions}
\label{sec:conclusions}
\setcounter{equation}{0}


Following the lack of missing-energy signals at the LHC, there is now
additional motivation to depart from the MSSM and
study in more detail alternative realisations of supersymmetry. A
promising scenario in this respect is provided by R-parity violation,
where missing energy is substituted by multi-lepton or multi-jet
events. While especially the latter can often be hard to distinguish from the
SM backgrounds, there are channels where the signals can be
spectacular.  Further motivation for such schemes is
provided by the fact that R-parity violation can be perfectly
compatible with supersymmetric dark matter, with a gravitino LSP that
is stable on cosmological scales, and rapid decays of the remaining
superparticles  inside the detector, thus being detectable.

Within this framework, we previously studied neutralino
decays at the LHC, paying particular attention to the rich flavour
structure of the theory. Here, we extend these studies to
direct RPV chargino decays, which
can take place in scenarios with a wino or higgsino effective LSP, where the lightest chargino and
neutralino have a small mass difference $\Delta m$.
In order to investigate the properties of these scenarios, we employ a bayesian scan over the MSSM
parameter space, taking into account the
relevant constraints. We perform our main scan with logarithmic priors, checking that the central conclusions remain unchanged with linear priors. In addition to concrete experimental signatures
related to RPV chargino decays,
we also identify distinct differences between a
wino and a higgsino LSP. For instance the first is somewhat preferred in the region of
small  $\Delta m$, while the second has a larger RPC branching ratio to leptons, which, if
a long-lived chargino is discovered, can be used to probe the gaugino
sector of the theory.

For the RPV decays we find the following interesting possibilities:
\begin{itemize}
\item
For $LL\bar{E}$ operators, spectacular signatures, such as three
charged-lepton resonances, with explicit lepton number
violation in the final state, can be expected. We have shown that this
can be the case for substantial patches of the supersymmetric
parameter space, even in the MSSM, which is the most
conservative scenario. In extensions of the theory, the
parameter space where such signatures could arise would be enhanced.
In addition, similarly to neutralino decays, there are
interesting features in the di-lepton invariant mass distributions,
where employing same-sign subtraction practically removes
the combinatorial background, thus revealing features that would otherwise
be invisible.
\item
For $LQ\bar{D}$ operators, the expected physics is similar to RPV neutralino decays, but with some quite
interesting differences. Particularly for $LQ_3\bar{D}$, there is a distinct difference between
RPV neutralino and chargino decays, since the first involves
neutrinos plus jets or a charged lepton and a top quark, while the
latter involves charged leptons plus jets or neutrinos and top quarks.
For charginos therefore, a signal with both a lepton and a b-tag and
potential charge identification has enhanced detectability.
\item
For $\bar{U}\bar{D}\bar{D}$ operators we do not expect significant chargino RPV decays for positive $\Delta m$. For small (below pion mass) and negative mass differences charginos below 500 GeV seem excluded by searches for long-lived charged particles. For heavier charginos detection through RPV decays seems very difficult, the use of sophisticated jet clustering methods is in general required. However, we note that the operators $\bar{U}_3 \bar{D}_j\bar{D}_k$ have a particularly interesting behaviour: while neutralino RPV decays will always contain a top quark, the chargino can decay to three jets including at least one b-jet. Furthermore, for the operators $\bar{U}_3 \bar{D}_j\bar{D}_3$ one can also get $bbj$ and $ttj$ final states, opening the possibility to search for rare  multi-top events and events with same sign top pairs.
\end{itemize}

Overall, signals from direct RPV chargino decays complement
previous studies, and looking for them is
necessary in order to make a complete search for supersymmetry at the
LHC. Moreover, since these signatures are directly correlated to the
flavour and group structure of the theory, which determines the particle mass
correlations and level of unification at high scales, it is hoped that
any observable signal will help to distinguish between different possible
realisations of supersymmetry.

\subsubsection*{Acknowledgements}
The CPU intensive parts of this work was performed on the Abel Cluster, owned by the University of Oslo
and the Norwegian metacenter for High Performance Computing (NOTUR), and operated by the Research Computing Services group at USIT, the University of Oslo IT-department. The computing time was given by NOTUR allocation NN9284K, financed through the Research Council of Norway. N.-E. Bomark is funded in part by the Welcome Programme of the Foundation for Polish Science. The use of the CIS computer cluster at the National Centre for Nuclear Research is gratefully acknowledged.


\end{document}